# Mathematical Modeling Analysis and Optimization of Fungal Diversity Growth

**Tongyue Shi**
School of Computer Science and Technology
Soochow University
Suzhou, China
tyshi@stu.suda.edu.cn

**Haining Wang**
School of Mathematical Sciences
Soochow University
Suzhou, China
hnwang@stu.suda.edu.cn

**Summary**

This paper studied the relationship between the decomposition rate of fungi and temperature, humidity, fungus elongation, moisture tolerance and fungus density in a given volume in the presence of a variety of fungi, and established a series of models to describe the decomposition of fungi in different states. Since the volume of soil was given in this case, the latter two characteristics could be attributed to the influence of the number of fungal population on the decomposition rate. Based on the Logistic model, the relationship between the number of population and time was established, and finally the number of fungi in the steady state was obtained The interaction between different species of fungi was analyzed by Lotka-Volterra model, and the decomposition rate of various fungal combinations in different environments was obtained.

Through the analysis of single fungal population and multiple fungal population, we made a corresponding comparison between the two situations. For the situation of multiple fungi is more complex, and the properties of various fungi are different, we considered the coexistence of the two fungal communities. After studying the one and two cases, we can extrapher from one to the other, and the community consisting of *n* fungal populations will be similar to the community consisting of *n+1* fungal populations.

After the study, we substituted the collected data into the model and found that the fungal community composed of two kinds of fungi had a lower decomposition rate of ground decomposition or wooden fiber than that of a single kind of fungus for the same kind of substance. After careful study, it was found that there might be some interaction between fungi, and the disturbance and competition among the populations had a certain influence on decomposition rate. So ecological diversity of fungi does not necessarily increase the rate of fungal decomposition.Then our decomposition rate of fungi in different environment are analyzed, in arid, semi-arid, temperate, the arbor and tropical rainforest region studied the data, we found that the fungus in warm and humid environment of decomposition rate is highest, the change of the atmospheric cause some fungal population growth rate decreases, there are also some will increase, which is associated with the nature of fungi.We analyzed the influence of environmental factors, namely temperature and humidity, on the model. Finally, we write an article on how biodiversity affects the efficiency of system decomposition and other conclusions, which is suitable for inclusion in an accepted university biology textbook.

# 1. Introduction

**1.1 A brief introduction to fungi**

Fungi are members of a group of eukaryotes, which include microorganisms such as yeasts, molds and the more familiar mushrooms. These organisms are classified as a kingdom, separate from the other eukaryotic kingdom of plants and animals. Fungi, as the third type of microorganisms in soil, are widely distributed in soil surface. Fungi in soil include algal fungi, ascomycetes, basidiomycetes and semi-known fungi, among which semi-known fungi are the most. Per gram of soil can contain $10^4 \sim 10^5$ fungi. In fungi, the mycelium of mold, like actinomycetes, develops and winds around the surface of organic debris and soil particles, spreads in the pores of the soil, and forms sexual or asexual spores.[1]

For soil molds are aerobic microorganisms and acid tolerant, they are generally distributed in the surface layer of soil and seldom develop in the deep layer. In the soil at pH5.0, soil fungi account for a higher proportion of the total soil microorganisms because the development of bacteria and actinomycetes is limited. Fungal hyphae are several to dozens of times wider than actinomycetes, so the biomass of soil fungi is not less than that of bacteria or actinomycetes. It is estimated that the length of fungal hyphae can reach 40m per gram of soil, and the live weight of fungi is about 0.6mg per gram of soil based on the average diameter of 5 m. The content of yeast in soil is small, about $10 \sim 10^3$ per gram of soil yeast, but the content in orchard, beekeeping soil is higher, per gram of orchard soil can contain $10^5$ yeast.

Due to the availability of nutrients, moisture, air, pH, osmotic pressure and temperature for the growth and development of various microorganisms, soil has become a good environment for microorganisms to live in. It can be said that soil is not only the "natural culture medium" of microorganisms, but also their "base camp". For classification, soil is the most abundant resource pool of bacteria. Generally speaking, the ratio of various microorganisms in each gram of cultivated soil has a series of decreasing rules:

bacteria ($\sim 10^8$) > actinomycetes ($\sim 10^7$, spores) > mold ($\sim 10^6$, spores) > yeast ($\sim 10^5$) > algae ($\sim 10^4$) > protozoa.

It can be seen that there are a large number of microorganisms in the soil, most of which are bacteria. It is estimated that there are about 150kg of mould, 75kg of bacteria, 15kg of protozoa, 7.5kg of algae and 7.5kg of yeast in each mu of cultivated soil. Through the vigorous metabolic activities of these microorganisms, the physical structure and fertility of soil can be significantly improved.

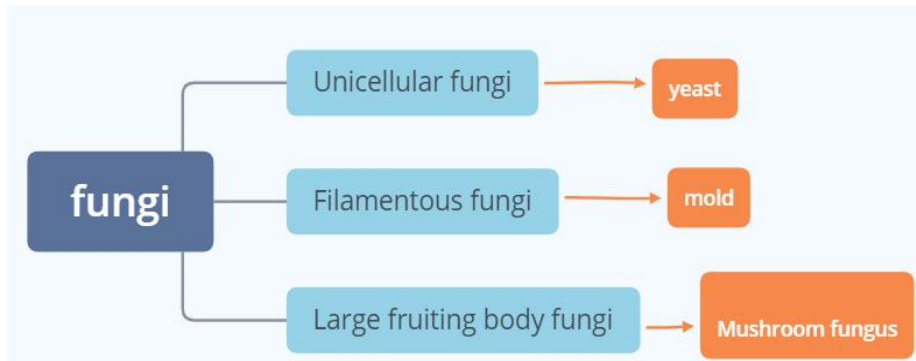

figure 1 Classification of fungi[2]

## 1.2 Brief introduction of carbon cycle and decomposition

Carbon cycle is a biogeochemical cycle of carbon exchange among the earth's biosphere, soil sphere, earth rock sphere, hydrosphere and atmosphere. Carbon is the main component of biological compounds, but also limestone and many other minerals. Like nitrogen cycle and water cycle, carbon cycle contains a series of events, which are the key to the earth's life support. It describes the movement of carbon in the cycle and reuse of the whole biosphere, as well as the long-term storage and release process of carbon sink and carbon sink. At present, land and ocean carbon sinks account for about a quarter of human carbon emissions each year.[3]

## 1.3 The decomposition of fungi

As the main organic matter in nature, lignocellulosic fiber constitutes 1 / 3 of annual plant weight and 1 / 2 of perennial plant weight, and plays an important role in carbon cycle, as shown in the figure. The main structure form of lignocellulose in compost is about 40%, hemicellulose about 20% - 30%, lignin about 20% - 30%. Although many microorganisms can decompose the isolated cellulose, the cellulose is protected by lignin in the cell wall, and lignin has a complete and hard shell, which is not easy to be degraded by microorganisms. Therefore, the decomposition of cellulose is limited.[4]

Thermophilic fungi have strong decomposition effect on cellulose, hemicellulose and lignin. They can not only secrete extracellular enzymes, but also have mechanical interpenetration effect on their hyphae. They can degrade refractory organic compounds (such as cellulose and lignin) in compost together, promote biochemical action, and play an important role in composting.

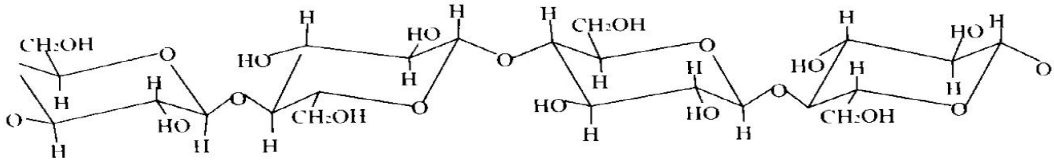

figure 2　Schematic diagram of cellulose structure

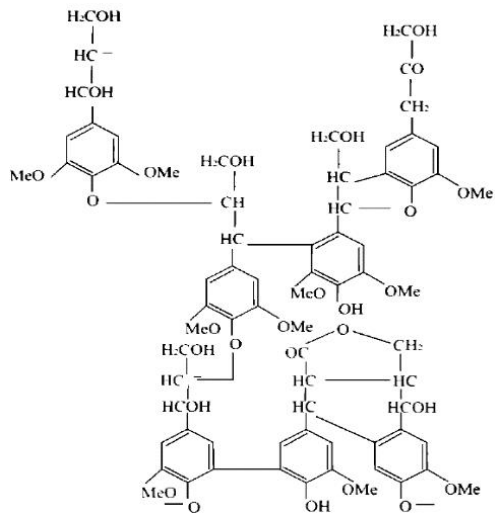

figure 3 Lignin structure

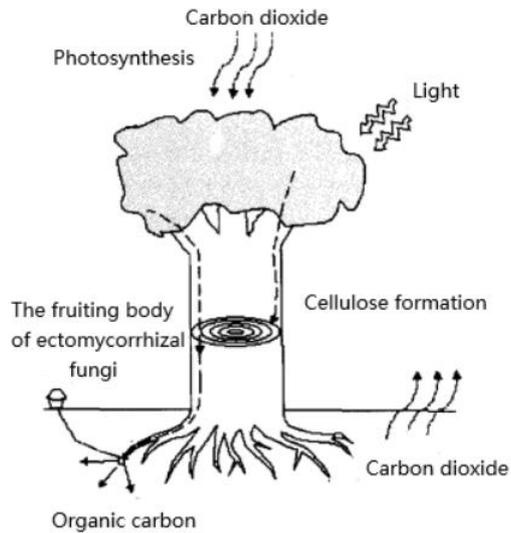

figure 4 Carbon cycle diagram[5]

**1.4 Mapping of mind**

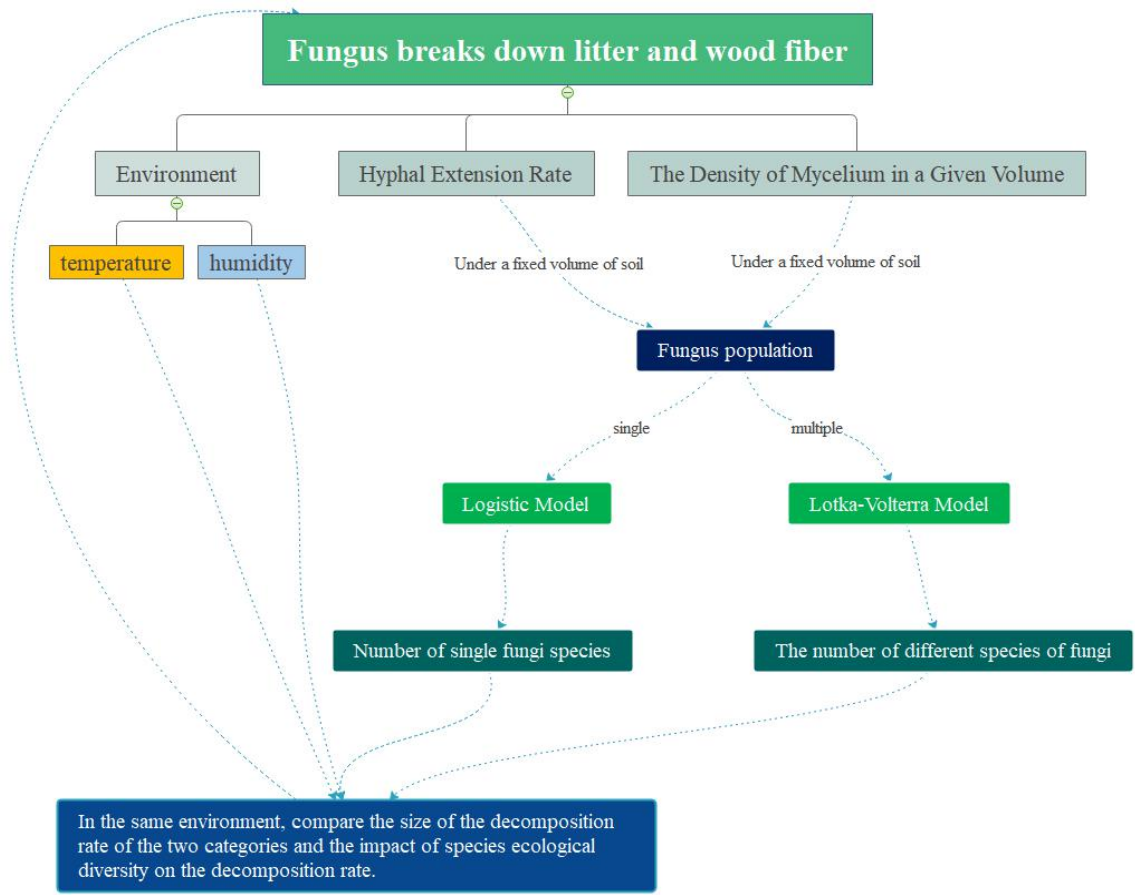

figure 5 Mapping of mind

## 2. Problem statement and analysis

### 2.1 Problem statement

1. Establish a mathematical model to describe the decomposition of litter and lignocellulosic fibers in the presence of multiple fungi.

2.In your model, incorporate interactions between different fungal species with different growth rates and different moisture tolerance, as shown in Figures 1 and 2.

provide an analysis of the model and describe the interactions between different types of fungi. The dynamics of the interactions should be described, including short-term and long-term trends. Your analysis should examine the sensitivity of the environment to rapid fluctuations and should determine the overall impact of changing atmospheric trends to assess the impact of changes in local weather patterns.

3.Including the prediction of the relative advantages and disadvantages of each species and possible persistent species combination, and the prediction of different environments such as arid, semi-arid, temperate, trees and tropical rain forests.

4.Describe how the diversity of fungal communities in the system affects the overall efficiency of the ground waste classification system. The importance and role of biodiversity prediction in local environment with varying degrees of variability. Attach a two page article to the results. Your article should be suitable for inclusion in University entry-level biology textbooks to discuss the latest progress in our understanding of the role of fungi in ecosystems.

## 2.2 Overall analysis

To solve this problem, we only need to establish a "decomposition rate model". Although "in the case of the existence of multiple fungi" is mentioned in the question, the focus of the original English sentence is "describe the breakdown of ground litter and wood fibers through fungal activity in the presence of multiple species of fungi" "Fungi" refers to the overall use of a method to describe the impact of fungi on plants. Therefore, the general result of this problem is a linear model of fungi decomposition rate. With the help of the background of the topic, the independent variable can be set as "temperature, humidity, overlapping degree between communities, time" and so on, and the dependent variable can be set as "decomposition rate (or vegetation decay rate / loss rate)" and so on. The data needed in the process of fitting the equation can be provided according to the topic, or other data can be found by oneself. Because it is the first question, we can consider as many factors as possible.

Question 2: in your model, combine the interactions of different kinds of fungi with different growth rates and different humidity tolerance. In particular, fungi with slow growth are more likely to survive and grow in the environment of humidity and temperature changes, while fungi with fast growth are less resistant to the same environmental changes. The "rate of change" of temperature and humidity is related to the resistance / vitality of fungi. The previous part mentioned "the relationship between growth rate and humidity tolerance", so this question is mainly modeled in two aspects: the expression of humidity tolerance and growth rate, and the combination of different fungi. When dealing with the relationship between humidity tolerance and growth rate, we can find that the relationship between humidity tolerance and growth rate of different fungi is approximately linear (as shown in the picture provided by the title), so we need to abstract it into a linear relationship, for example. Therefore, the combination of fungi with different growth rate and humidity tolerance means to combine the above two factors while keeping the "decomposition effect (decomposition rate)" of fungi unchanged. Because their combined effect is the decomposition rate, so as long as the decomposition rate remains unchanged, and then the two factors are combined (at the same time, this requires the help of the "growth rate and humidity tolerance relationship")。
In this way, we can get the combined "fungal decomposition rate" model.

Question 3: provide a model to describe the interaction between different types of fungi. The dynamic description of interaction should include short-term and long-term trends. Your analysis should also examine the sensitivity to fluctuations in environmental changes and determine the constant trends of the atmospheric environment to assist in assessing the impact of weather changes on the model. Because different fungi have different humidity tolerance, their growth rates are different in the same environment, which leads to the interaction between different types of fungi. Through the question, we need to see another key point, that is, why do fungi interact with each other? As a matter of fact, the influence of fungal communities mentioned in the previous article, generally speaking, is the competitive relationship between populations, which is mainly reflected in "whether there is nutrition supply" and "whether there is oxygen". So it's easy to understand. The model we need to build is: in the same initial environment, what type of community evolution will happen between different types of fungi. The fungi that are more suitable for the environment temperature and humidity will grow faster. Although two or more kinds of fungi will not compete under the condition of sufficient nutrition supply at the beginning, when the number of fungi that are more suitable for the environment is far greater than that of the vulnerable side, the vulnerable fungi will lack nutrition and may lead to the death of fungi.

If you want to further deepen the topic, you can further consider a relationship, that is, the influence of "whether there is oxygen" on aerobic respiration and anaerobic respiration. If the number of strong fungi is too much, will it affect the oxygen of weak fungi? This point can be further studied by referring to the literature. At the same time, assuming that the influence of oxygen is ignored, if both sides are engaged in aerobic respiration or anaerobic respiration or mixed aerobic and anaerobic respiration, will the products of these two kinds of respiration further affect the survival of each other's fungi? For example, will the alcohol produced inhibit the growth of the other side's fungi, or inhibit the growth of its own side? Therefore, through the above analysis, we can roughly get the short-term and long-term trend of the interaction required by this topic. As for the sensitivity to environmental fluctuations, it means whether the small changes in initial conditions will affect the different trends of subsequent community evolution? Will it cause the original strong fungi to become weak fungi? After considering the influence of the environment on the initial conditions, the initial environmental factors can be added to the model to further assist the influence of weather changes on the model.

Question 4: it also includes the prediction of the relative advantages and disadvantages of "each species" or "species combination that may last for a period of time", and the prediction of different environments such as arid, semi-arid, temperate, arbor, tropical rain forest, etc. This question is similar to the thinking direction of question 3, mainly considering "dynamism", that is, a trend. For the "prediction of relative advantages and disadvantages", we can use time as the measurement standard, and then judge the relative advantages and disadvantages of fungi in the change; for the "drought and humidity" and other environmental prediction, that is, the impact of initial environmental conditions on community evolution. The factors that can be considered are: the duration of temperature and humidity, and what kind of fungi may exist in arid and humid areas.

Question 5: describe how the diversity of fungal communities affects the overall decomposition efficiency of litter. When the local environment has different degrees of variability, it is important to predict the importance and role of biodiversity. This question is based on the number of fungi as an independent variable to consider, just need to design a model to analyze the impact of the number of fungi on the decomposition efficiency. For example, you can use the "community evolution model" in question 3, set the initial condition to the number of fungi, and then see how long the litter will be decomposed. When considering the local environment may have different degrees of variability, that is, different initial conditions, the "anti-interference ability" and "resilience" of diverse fungal communities are stronger, which should be easier to think of. The effect of fungi species on decomposition efficiency can be added into the judgment index, and the conditions under which the fungi species diversity can be completely restored (as far as possible) to the original state can be added.

## 3. Basic assumption

- Assuming that the environmental capacity of the two fungi in the model is the environmental capacity of the fungus under the most suitable temperature and humidity conditions.
- Assuming a linear relationship between fungal decomposition rate and humidity.
- Assuming a linear relationship between the decomposition rate of fungi and the number of fungi.

## 4. Symbols

Table.1 Variables and their meanings

| Number | Sign | Significance |
|---|---|---|
| 1 | $D$ | Fungal decomposition rate |
| 2 | $T$ | Ambient temperature |
| 3 | $H$ | Environment humidity |
| 4 | $N$ | Fungus population |
| 5 | $K$ | Maximum environmental capacity of fungus population |
| 6 | $a, b$ | Population competition coefficient |

# 5. Model

**5.1 Factors affecting fungal decomposition rate**

According to the search literature, charcoal is decomposed by fungi into glucose, carbon dioxide and water. It is concluded that the decomposition rate of fungi is proportional to the emission of carbon dioxide. The study by Buyanoyskyt shows that the intensity of soil respiration, that is, the rate of carbon dioxide emission is proportional to temperature. , The increase in temperature promotes the decomposition of organic carbon in the soil and accelerates the biological turnover of soil microorganisms. Then we can conclude that the decomposition rate of fungi is also proportional to temperature.

Soil moisture affects soil physical and chemical properties and soil microbial characteristics. The decomposition rate of organic carbon is still different for different soil moisture content. We know from the literature that under soil moisture conditions (30%-90% field water holding capacity) ), the soil carbon dioxide emission rate increases with the increase of soil moisture content, so we can assume that the decomposition rate of fungi has a linear relationship with humidity.

Under laboratory conditions, fungal growth rate (hyphae elongation rate) is the strongest single predictor of fungal-mediated wood decomposition rate, and accounts for 27% of in-situ variation in field decomposition. Mycelial elongation rate is fungus The growth rate of the hyphae is a very important feature of fungi. Hyphae refer to the cells that stretch out and form the filaments and structures of the fungus. Different types of hyphae play different roles in the life cycle of fungi. Another feature is the density of mycelium in a given volume, because the volume of soil we assume is fixed. Therefore, the density of the mycelium in a given volume is the number of the fungal population on the patch, and the effect of the mycelial elongation rate on the decomposition rate is essentially the effect of the population size on the decomposition rate. Mycelial elongation and mycelial density reflect mycelial morphology and growth strategy, but in fact the decomposition process of different types of fungi is different. In nature, there are only a few organisms that can degrade lignin and produce corresponding enzymes. The complete degradation of lignin is the result of the joint action of fungi, bacteria and corresponding microbial communities. Among them, fungi play a major role. Fungi that degrade lignin It can be divided into white rot fungi, brown rot fungi and soft rot fungi according to their decay conditions. Their decomposition rates are different. Simply put, the fungi contain different lignin degrading enzymes and the number of degradable enzymes It's not the same. According to the research, there are three most important lignin degrading enzymes, namely lignin peroxidase, manganese-dependent peroxidase and laccase. However, the types of enzymes produced by fungi and some physical and chemical properties are different, so it is assumed there is a linear relationship between the decomposition rate of fungi and the number of fungi, and the coefficient before the number of fungi depends on the nature of the fungus itself. Therefore, different types of fungi , In the relationship between the decomposition rate of fungi and the number of fungi, the current coefficient of the number of fungi will also vary. In different environments, the humidity resistance of various fungi and their adaptation to temperature are different, so when we set up the equation, the humidity of various fungi and the coefficient before humidity will also change due to different fungi. We suppose that the decomposition rate of fungi is D, the temperature is T, and the absolute humidity is H. The number of a certain fungus population is N, then we can set up an equation:

$$D = a_1T + a_2H + a_3N + a_4$$

Table 2 : Hyphal extension rate measured for each isolate
under standardized laboratory conditions at 10, 16 and 22 °C(under relative humidity 70%)

|  | Extension rate (/mm day) ± SD | | |
|---|---|---|---|
| Isolate | 10℃ | 16℃ | 22℃ |
| *Armillaria gallica* FP102531 C6D (south) * | 0.30 ± 0.05 | 0.36 ± 0.05 | 0.34 ± 0.06 |
| *Armillaria gallica* EL8 A6F (north) * | 0.18 ± 0.06 | 0.26 ± 0.05 | 0.38 ± 0.15 |
| *Armillaria gallica* FP102534 A5A (south) * | 0.26 ± 0.05 | 0.24 ± 0.05 | 0.32 ± 0.06 |
| *Fomes fomentarius* TJV93 7 A3E | 0.36 ± 0.08 | 1.28 ± 0.22 | 4.62 ± 0.24 |
| *Hyphodontia crustosa* HHB13392 B7B | 1.20 ± 0.03 | 0.99 ± 0.07 | 1.77 ± 0.20 |
| *Pycnoporus sanguineus* PR SC 95 A11C | 0.81 ± 0.03 | 3.21 ± 0.07 | 7.26 ± 0.17 |
| *Schizophyllum commune* TJV93 5 A10A | 1.88 ± 0.25 | 3.32 ± 0.08 | 7.40 ± 0 |
| *Schizophyllum commune* PR1117 | 1.06 ± 0.04 | 1.64 ± 0 | 4.60 ± 0.41 |
| *Tyromyces chioneus* HHB11933 B10F | 1.92 ± 0.13 | 3.37 ± 0.10 | 5.67 ± 0.12 |
| *Xylobolus subpileatus* FP102567 A11A | 0.74 ± 0.10 | 1.00 ± 0.09 | 1.04 ± 0.09 |
| …… | …… | …… | …… |

Table 3: Hyphal extension rate measured for each isolate
under standardized laboratory conditions at different relative humidity(under 16℃)

|  | Extension rate (/mm day) ± SD | | |
|---|---|---|---|
| Isolate | 60% | 70% | 80% |
| *Armillaria gallica* FP102531 C6D (south) * | 0.23 ± 0.03 | 0.36 ± 0.05 | 0.25 ± 0.03 |
| *Armillaria gallica* EL8 A6F (north) * | 0.16 ± 0.02 | 0.26 ± 0.05 | 0.33 ± 0.15 |
| *Armillaria gallica* FP102534 A5A (south) * | 0.21 ± 0.02 | 0.24 ± 0.05 | 0.28 ± 0.06 |
| *Fomes fomentarius* TJV93 7 A3E | 0.36 ± 0.07 | 1.28 ± 0.22 | 2.62 ± 0.23 |
| *Hyphodontia crustosa* HHB13392 B7B | 1.10 ± 0.04 | 0.99 ± 0.07 | 1.43 ± 0.14 |
| *Pycnoporus sanguineus* PR SC 95 A11C | 0.75 ± 0.02 | 3.21 ± 0.07 | 6.26 ± 0.23 |
| *Schizophyllum commune* TJV93 5 A10A | 1.43 ± 0.25 | 3.32 ± 0.08 | 6.54 ± 0.05 |
| *Schizophyllum commune* PR1117 | 1.09 ± 0.05 | 1.64 ± 0 | 3.98 ± 0.34 |
| *Tyromyces chioneus* HHB11933 B10F | 1.98 ± 0.14 | 3.37 ± 0.10 | 4.57 ± 0.02 |
| *Xylobolus subpileatus* FP102567 A11A | 0.75 ± 0.11 | 1.00 ± 0.09 | 1.34 ± 0.07 |
| …… | …… | …… | …… |

**5.2 Logistic model**

① There is an environmental capacity (usually expressed in $K$), when $N = K$ the population growth is zero, that is $dN/dt = 0$.

② The decrease of growth rate with the increase of density is proportional. The simplest $1/K$ is the inhibitory effect of each additional individual. In other words, suppose that a certain space can only accommodate $K$ individuals, the $1/K$ space that each individual has used, the $N$ individual has used the $N/K$ space, and the "remaining space" that can provide for the continuous growth of the population is only $(1-N/K)$. According to these two hypotheses, the density constraint leads to the decrease of T with the increase of density, which is contrary to the situation of non density constraint. The population growth is no longer a "t" shape, but a low "s" shape. The "s" curve has two characteristics:

① The curve is asymptotically close to the value, that is, the equilibrium density.

② Curve rising is smooth. The simplest mathematical model for generating "s" curve can be explained and described as the above exponential growth equation multiplied by a density constraint factor to obtain the famous logistic equation in the history of ecological development.

$$dN/dt = rN(1-N/K)$$

The integral formula is as follows:

$$N_t = K/(1+e^{a-n})$$

Where the value of parameter a depends on the N. It represents the relative position of the curve to the origin.

In the early stage of population growth, the population size n is very small, and the N / K value is also very small, so (1-N / K) is close to 1. Therefore, the inhibition effect can be ignored, and the population growth is essentially rN, showing geometric growth. However, when n is larger, the inhibition effect increases, until when n = k, (1-N / k) becomes (1-k / R), equal to 0, then the population growth is zero, and the population reaches a stable size invariant equilibrium.

The two parameters r and K in logistic equation have important biological significance. r is the population growth capacity, K is the environmental capacity, that is, the equilibrium density of species in a specific environment. However, it should be noted that K, like other ecological characteristics, also changes with the change of environmental conditions (resources). In addition, the reciprocal of instantaneous growth rate $T_r$ is also a useful parameter, which is called natural reaction time. The larger r is, the faster the population growth is, and the smaller r is, which means that the time needed for the population to return to the equilibrium state after being disturbed is shorter, and vice versa. $T_r$ is an important parameter to measure the time required for the population to return to the normal state after being disturbed.

There are also interactions among different fungal populations. Based on the competition experiment of Paramecium, cause put forward Gauss hypothesis, which was developed into competition exclusion principle. The contents are as follows: in a stable environment, two or more species limited by resources but with the same way of resource utilization cannot co-exist for a long time, that is, complete competitors cannot co-exist.

There are two ways of competition, or through the loss of limited resources .However,

individuals do not interact with each other directly, or compete through the direct interaction between competing individuals (interference competition).

The interaction between the two fungi can be simply assumed as intraspecific competition, which can be simulated by Lotka-Volterra model.

x=[0.5 1 1.5 2 2.5 3 3.5 4 4.5 5 ];

z=[0 1.90 2.80 4.7171 6.622 8.97666 11.21 11.34 11.41 11.43]

y=[1 80 640 52135 4192000 947680000 162520000000 222520000000 259632000000 275200000000];

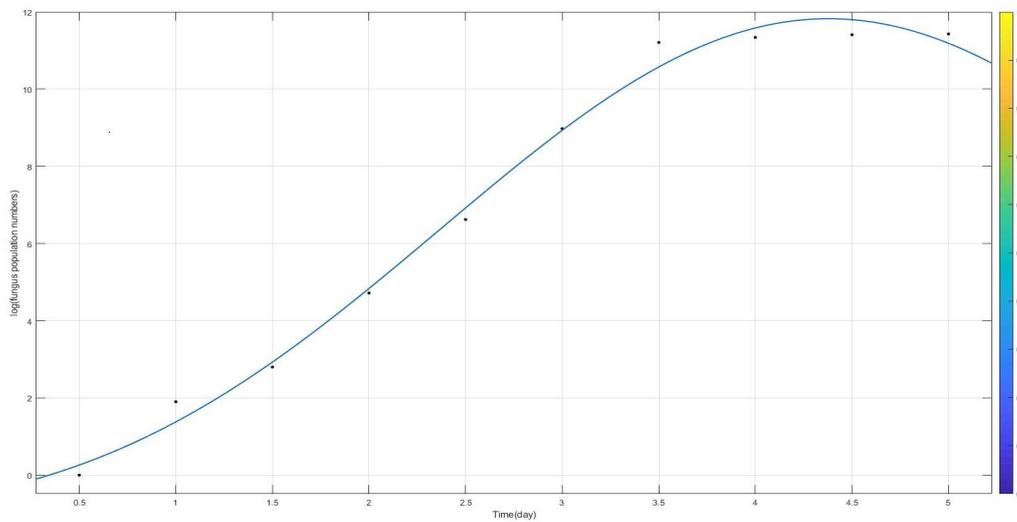

figure 6 Growth curve of fungus1

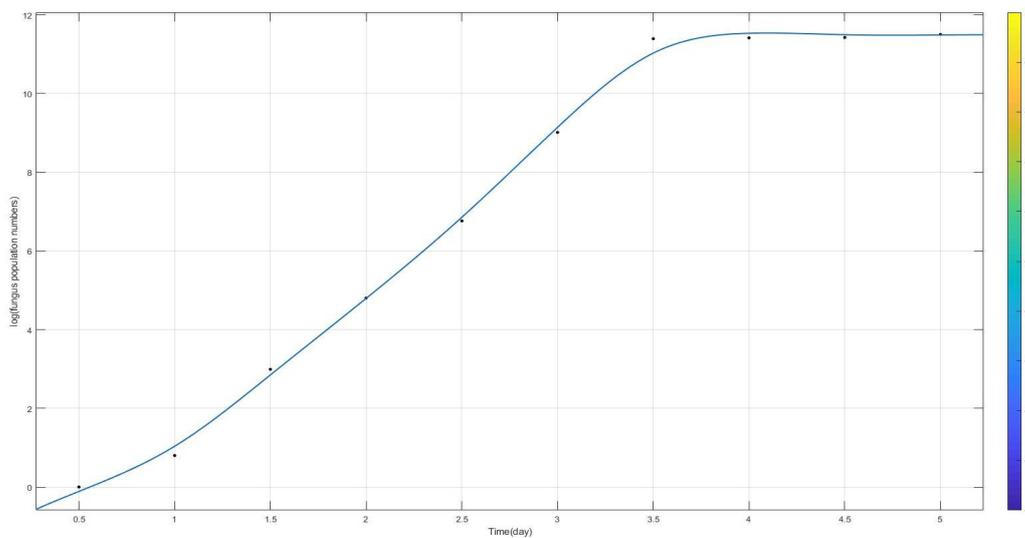

figure 7 Growth curve of fungus2

## 5.3 Lotka-Vlterra model

Lotka-vlterra's interspecific competition model is an extension of logistic model. Let sum $N_1$ and $N_2$ be the population number of the two species respectively, and $K_1$ $K_2$ $r_1$ $r_2$ be the environmental capacity and population growth rate of the two species. According to the logistic model[6]:

$$dN_1/dt = r_1 N_1 (1 - N_1/K_1)$$

As mentioned earlier, $(1 - N/K)$ item can be understood as an unused "remaining space" item, but $N/K$ is a "utilized space item". When two species compete or use space together, the used space items should be added in $N_2$ addition to $N_1$, that is:

$$dN_1/dt = r_1 N_1 (1 - N_1/K_1 - aN_2/K_1) \qquad (1)$$

$a$ is the competition coefficient, which means that the space occupied by $N_2$ individuals is equal to $a$ of $N_1$ individual. For the competition inhibition effect. Similarly, for species 2:

$$dN_2/dt = r_2 N_2 (1 - N_2/K_2 - bN_1/K_2) \qquad (2)$$

β is the competition coefficient of species 1 to species 2. The sum of the equations (1) and (2) are the competition model of lolka Volterra.

Based on Lotka-vlterra model, the relationship between the number of fungi and time was established Lotka-vlterra's interspecific competition model is an extension of logistic model. Because the environmental capacity of each fungus is different in various environments, the changes of temperature and humidity will also affect the environmental capacity of the population. Because the relationship between the environmental capacity of different fungal populations and temperature and humidity is very complex, and the rules between them are also very different, so we assume that the environmental capacity of two fungi in the model is different The environmental capacity is the capacity of the fungus under the most suitable temperature and humidity conditions, that is, the absolute maximum environmental capacity of the fungus.

$$r_1(N_1) = r_1 - SN_1'$$

The number of fungi that can be accommodated by natural resources and environmental conditions is as followd。 From the assumption we can get:

$$r_1(N_1) = r_1(1 - N/K_1)$$

After calculating the differential equation, we can get:

$$\frac{dN_1}{dt} = 0$$

$$\frac{dN_1}{dt} = r_1(1 - \frac{N_1}{K_1})N_1$$

$$x(t_0) = x_0$$

Solve for:

$$N_1(t) = \frac{K_1}{1 + (\frac{K_1}{x_0} - 1)e^{-n(t-t_0)}}$$

In the model, there are three competing outcomes of the two fungi

1. Fungi 1 was better and fungi 2 was excluded

2. Fungi 2 was better than fungi 1 was excluded

3. When the two coexist, they are in equilibrium

However, the final result of case 0.102 is the case of a single fungus, so the first and second cases will not be considered, so we will consider the coexistence of the two cases

Through the information collected from the database, we can get the data of two carts of fungi co existing in one cubic meter of soil. We can get the difference from the knowledge in ordinary differential equation. Because the other side of differential equation is a continuous function, so it has only one solution. Then we can input the data in the data in the code, and we can get two solutions through data input The distribution of fungi in soil at steady state.

Finally, we substitute the n of the last steady state in one dimension into the formula to find it, and then combine the number of two kinds of fungi in the second case to find the decomposition rate D of the two kinds of fungi. We list the following figure.

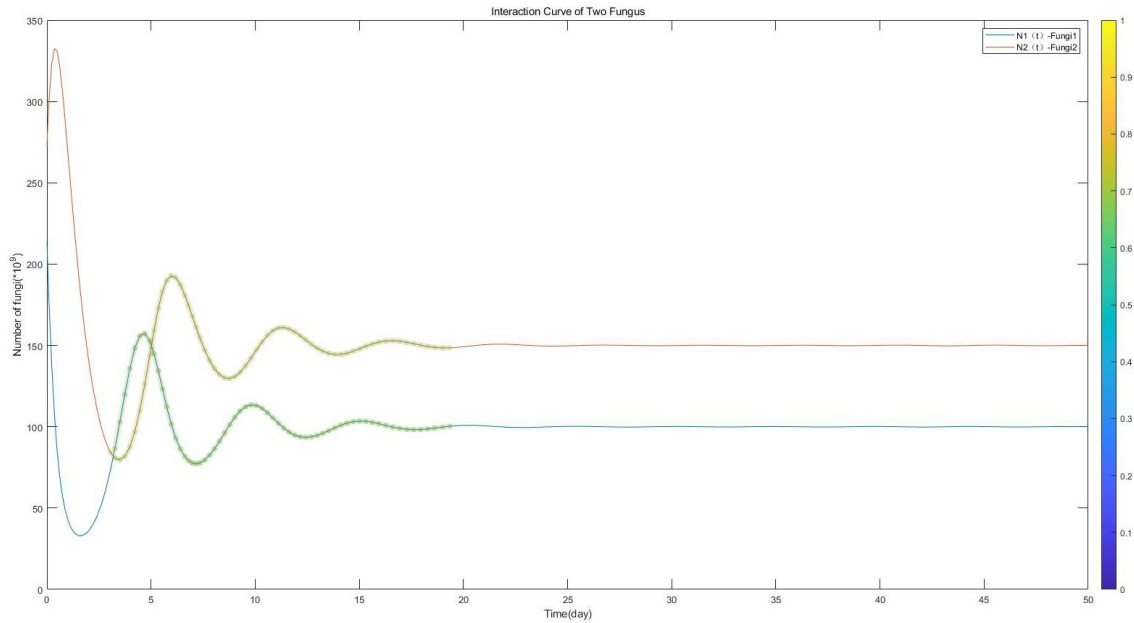

figure 8 Growth curve of fungus2

A terrestrial ecosystem is a type of ecosystem found only on land forms. Six primary terrestrial ecosystems exist: tundra, taiga, temperate deciduous forest, tropical rain forest, grassland, deserts.

For different environments including arid, semi-arid, temperate, arboreal, and tropical rain forests. Here, it can be assumed that arid corresponds to desert climate, semi-arid corresponds to grassland climate, temperate corresponds to temperate deciduous broad-leaved forest, arboreal corresponds to subtropical evergreen broad-leaved forest, and finally tropical rain forest. With reference to some data, the main characteristics of different terrestrial ecosystems in different climate environments are obtained, and the main temperature (annual average temperature), humidity (annual precipitation), distribution location, etc. are listed, see the figure below.

The analysis of decomposition rate under different climate conditions is shown in the **conclusion** section.

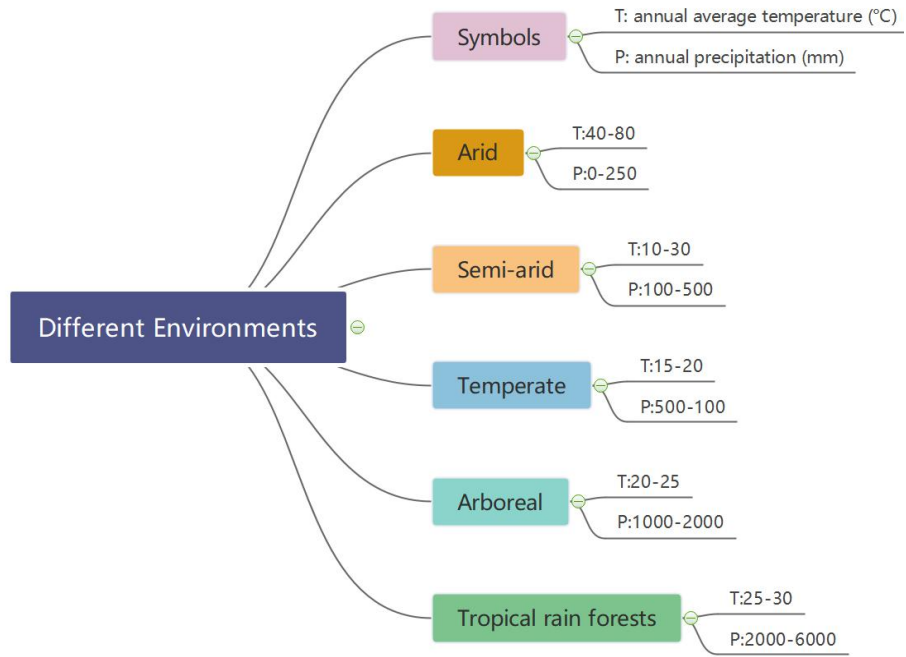

figure 9 Different climate environments[7]

## 6 Sensitivity Analysis

In our model, environmental factors measure the effect on the decomposition rate of fungi. We sorted the same fungi under different temperatures and different humidity one by one, and got the following chart:

| Table 4: Decomposition rate and temperature under single and interaction | | | |
|---|---|---|---|
| **Single** | 10℃ | 16℃ | 22℃ |
| Armillaria gallica | 0.30 ± 0.05 | 0.36 ± 0.05 | 0.34 ± 0.06 |
| Fomes fomentarius | 0.36 ± 0.08 | 1.28 ± 0.22 | 4.62 ± 0.24 |
| **Interaction** | 0.28 ± 0.03 | 0.66 ± 0.03 | 2.56 ± 0.05 |

| Table 5: Decomposition rate and relative humidity under single and interaction | | | |
|---|---|---|---|
| **Single** | 60% | 70% | 80% |
| Armillaria gallica | 0.23 ± 0.03 | 0.36 ± 0.05 | 0.25 ± 0.03 |
| Fomes fomentarius | 0.36 ± 0.07 | 1.28 ± 0.22 | 2.62 ± 0.23 |
| **Interaction** | 0.22 ± 0.02 | 0.42 ± 0.02 | 1.23 ± 0.02 |

We change the corresponding temperature and humidity conditions to get the corresponding decomposition rate. From the table, we can see that as the environment becomes hotter and the humidity increases, the interval between their decomposition rates changes smaller. When the surface temperature and humidity are constant, environmental factors will still have a certain effect on the decomposition rate.

# 7 Conclusion

According to the results of the model, the decomposition rate of the combination of two fungi is lower than that of the combination of single fungi under the same environment. After careful exploration, it was found that the growth rate curve of single species population was S-shaped, possibly due to the interaction between fungi and the competition between populations.

At first, when the number of fungi was small, the growth rate of fungi increased, but the growth rate of different fungi was also affected by environmental changes, and various fungi had different tolerance to water. Some fungi have a lower growth rate, but the growth rate will change in response to changes in the environment. The decomposition rates between *n*-fungal combinations and *n+1*-fungal combinations can be compared to the decomposition rates of two fungal combinations and single fungal species. It can be found that the higher the fungal biodiversity, the slower the rate of decomposition. According to the conditions of arid, semi-arid, temperate, arboreal and tropical rain forests, for various fungal combinations, it was found that the higher the temperature and humidity were, the higher the decomposition rate of fungi would be. According to some studies, there are three species of plant parasitic macrofungi, all of which are dead parasitic fungi, and no obligate parasitic macrofungi have been reported.

There are different fungal combinations in different environments. Soil-born saprophytic fungi account for the largest proportion of the 200 species of macrofungi investigated in temperate regions, but the situation in this part of the fungi is more complex. Therefore, we should improve the profit distribution mechanism of farmland circulation, establish a relatively perfect farmland circulation market, and make rational use of chemical fertilizers and pesticides.

Fungi play an important role in the carbon cycle by decomposing organic matter into inorganic substances. There are a large number of fungi in the nature of rotten wood. Biodiversity is very important for the natural ecological environment. It can be seen from the model that fungal diversity has a very important ecological service function on a global scale by maintaining the niche of rotten wood and reducing the rate of $CO_2$ emission from rotten wood decomposition into the atmosphere. Therefore, fungal ecological diversity is of great benefit to the ecological environment.

Ecological diversity affects the overall efficiency of the ground waste decomposition and enables the stable progress of carbon cycle. It has a great impact on the smooth operation of ecology, so that the emission and absorption of carbon dioxide can be in a balanced state, and ecological diversity has direct, indirect and potential value. Change in the environment, and the fungus in the community, some fungi population growth rate will increase along with the change of environment. Some fungi population growth rate will decrease with the change of the environment, According to different fungal species, the growth rate of the fungal community will be more stable than single fungus species. Their moisture resistant also each are not identical, so the ecological diversity of fungal community makes fungus decomposed garbage ground efficiency more stable. When the environment changes, the fungal community will be relatively little affected, and all aspects of the situation is relatively stable, making the ecosystem can be stable.

# 8 Evaluation and Spread of the Model

In the world of microorganisms, we use knowledge of mathematics, biology, chemistry, and ecology to establish related models for the decomposition of fungi. The Lotka-Volterra model contains one-dimensional and two-dimensional models. Through analogy one and two models, we can get relevant conclusions about the decomposition of n species and n+1 species of fungal communities. We have known the decomposition rate of fungal communities in various environments, and conducted corresponding studies on various fungal combinations to further study the effects of ecological diversity. The system decomposes the impact of ground garbage, and concludes that ecological diversity is of great significance for stabilizing the ecological cycle.

Throughout the modeling process, we made reasonable assumptions and simplified the problem. We provided convenience for constructing and solving the model. Finally, analyze the sensitivity of the model and tested the stability of the model.

We transformed the problem of fungal decomposition rate into a problem of fungal community, greatly simplified the problem, and explored related influencing factors.

However, there are still many deficiencies in this model. There are many parameters in the model. These parameter values are related to biology. The corresponding parameters of different species of fungi are different, and there are many fungi, so we can only study representative ones. For fungi, the influence of parameters may bring corresponding changes to the results of the model.

In order to simplify the model, we assume that the growth rate of the fungal population has a linear relationship with time, but in fact there are environmental factors that will affect it, and environmental changes are not considered, so the model has certain limitations.